\documentclass[12pt]{article}
\usepackage[bibtex-style,alphabetic]{amsrefs}
\usepackage{graphicx}
\usepackage{amsmath, amssymb, amsthm}
\usepackage{dsfont}
\usepackage{setspace}
\usepackage{csquotes}
\usepackage{hyperref}
\usepackage[utf8x]{inputenc}
\usepackage[english]{babel}
\usepackage{hebfont}
\usepackage[toc,page]{appendix}
\usepackage{float}
\usepackage{enumitem}

\usepackage{anysize}

\usepackage{amsmath, amssymb}
\usepackage[percents]{overpic}

\usepackage{xspace}
\makeatletter

\DeclareRobustCommand\onedot{\futurelet\@let@token\@onedot}
\def\@onedot{\ifx\@let@token.\else.\null\fi\xspace}
 
\def\ie{\emph{i.e}\onedot}

\def\etal{\emph{et al}\onedot}

\marginsize{3cm}{3cm}{2cm}{1cm}

\begin{document}
\selectlanguage{english}
\pagestyle{empty}
\begin{center}
\vspace{5cm}
\doublespacing
\selectlanguage{english}
{\large \uppercase{\textbf {Matching LBO eigenspace of non-rigid shapes via high order statistics}}}
\vspace{5mm}

\vspace{2cm} {\Large Alon Shtern, Ron Kimmel} \\

\vspace{3cm}
{Technion - Israel Institute of Technology}\\
\end{center}

\newpage

\pagestyle{plain}
\pagenumbering{roman}
\newpage

\pagenumbering{arabic} \setcounter{page}{1}


\begin{abstract}
\label{sec:abstract}
A fundamental tool in shape analysis is the virtual embedding of the Riemannian manifold describing the geometry of a shape into Euclidean space.
Several methods have been proposed to embed isometric shapes in flat domains while preserving distances measured on the manifold.
Recently, attention has been given to embedding shapes into the eigenspace of the Lapalce-Beltrami operator.
The Laplace-Beltrami eigenspace preserves the diffusion distance, and is invariant under isometric transformations.
However, Laplace-Beltrami eigenfunctions computed independently for different shapes are often incompatible with each other.
Applications involving multiple shapes, such as pointwise correspondence, would greatly benefit if their respective eigenfunctions were somehow matched.
Here, we introduce a statistical approach for matching eigenfunctions.
We consider the values of the eigenfunctions over the manifold as sampling of random variables, and try to match their multivariate distributions.
Comparing distributions is done indirectly, using high order statistics.
We show that  the permutation and sign ambiguities of low order eigenfunctions, can be inferred by minimizing the difference of their third order moments.
The sign ambiguities of antisymmetric eigenfunctions can be resolved by exploiting isometric invariant relations between the gradients of the eigenfunctions and the surface normal.
We present experiments demonstrating the success of the proposed method applied to feature point correspondence.

\end{abstract}


\section{Introduction}
\label{sec:introduction}

The embedding of nonrigid shapes into a Euclidean space is well established, and widely used by shape analysis applications.
Usually, the mapping from the manifold to the Euclidean space preserves distances, that is the distance measured between two points on the manifold is approximated by the respective distance calculated in the Euclidean space.
The embedding of multiple isometric shapes into the same common Euclidean space seems to be ideal for applications like pointwise correspondence and shape editing.
A useful property of this common embedding would be if any corresponding points of different isometric shapes were mapped to nearby target points in the Euclidean space.
If this property is fulfilled then the simultaneous processing of shapes in the target domain can be done in a straightforward manner.

Elad \etal \cite{elad2003bending} used classical MDS embedding into the geodesic kernel eigenspace. 
The MDS dissimilarity measure was based on the geodesic distances computed by the {\em fast marching} procedure \cite{kimmel1998computing}. 
B{\'e}rard \etal \cite{berard1994embedding}  used the heat operator spectral decomposition to define a metric between two manifolds $M$ and $M'$.
They embedded the two manifolds into their respective eigenspaces.
They showed that the Gromov Haussdorf distance between the embedded manifolds $d_{GH}(M, M')=0$ if and only if the Riemannian manifolds $M$ and $M'$ are isometric.
Lafon \etal \cite{coifman2005geometric} defined the {\em diffusion maps} and showed that the embedding into the heat kernel eigenspace is isometry invariant, and preserves the diffusion metric.
Rustamov \cite{rustamov2007laplace} introduced the {\em Global Point Signature} (GPS) embedding for deformation invariant shape representation.

Although the diffusion maps computed independently on isometric shapes have a nearly compatible eigenbasis,  several inconsistencies arise:
\begin {itemize}
\item Eigenfunctions are defined up to a sign.
\item The order of the eigenfunctions, especially those representing higher frequencies, is not repeatable across shapes. 
\item The eigenvalues of the Laplace-Beltrami operator may have multiplicity greater than one, with several eigenfunctions corresponding to each such eigenvalue.
\item It is generally impossible to expect that an eigenfunction with large eigenvalue of one shape will correspond to {\em any} eigenfunction of another shape.
\item  Intrinsic symmetries introduce self-ambiguity, adding complexity to the sign estimation challenge.
\end{itemize}
These drawbacks limit the use of diffusion maps in simultaneous shape analysis and processing, they do not allow using high frequencies, and usually require some intervention to
order the eigenfunctions or solve sign ambiguities.

In this paper we present a novel method for matching eigenfunctions that were independently calculated for two nearly isometric shapes.
We rely on the fact that for low order eigenfunctions, inconsistencies are governed by a small number of discrete parameters characterized by the sign sequence and permutation vector. 
We estimate these parameters by matching statistical properties over the spectral domain.
The matching of the corresponding eigenfunctions enables the use of diffusion maps for consistent embedding of multiple isometric shapes into a common Euclidean space.

\subsection{Related Work}
The problems of eigenfunctions permutation and sign ambiguity were previously addressed in the context of simultaneous shape processing.
Several authors, among them Shapiro and Brady \cite{shapiro1992feature}, and Jain \etal \cite{jain2007non}, proposed using either exhaustive search or greedy approach for the eigenvalue ordering and sign detection.
Umeyama \cite{umeyama1988eigendecomposition} proposed using a combination of the absolute values of the eigenfunctions and an exhaustive search.
Mateus \etal \cite{mateus2007articulated} expressed the connection between the eigenfunctions of two shapes by an orthogonal matrix.
They formulated the matching as a global optimization problem, optimizing over the space of orthogonal matrices, and solved it using the expectation minimization approach. 
Later, Mateus \etal \cite{mateus2008articulated} and Knossow \etal \cite{knossow2009inexact} suggested using histograms of eigenfunctions values to detect their ordering and signs.
Dubrovina \etal \cite{dubrovina2011approximately}, suggested using a coarse matching based on absolute values of eigenfunctions together
with geodesic distances measured on the two shapes.

Most of these methods do not reliably resolve eigenfunction permutation \cite{jain2007non,knossow2009inexact,mateus2007articulated,shapiro1992feature}. 
Some of the above algorithms are limited by high complexity and do not allow the matching of more than a few eigenfunctions \cite{umeyama1988eigendecomposition,dubrovina2011approximately}.
None of these methods reliably estimate the sign sequence of antisymmetric eigenfunctions.
At the other end, Kovnatsky \etal \cite{kovnatsky2012coupled} proposed to avoid the matching problem by constructing a common approximate eigenbases for multiple
shapes using approximate joint diagonalization algorithms. Yet, it relies on a prior knowledge of a set of corresponding feature points.

Finally, the algorithm proposed by Pokrass \etal \cite{pokrass2012sparse} mostly resembles our approach. They used sparse modelling to match the LBO eigenfunctions that spans the {\em Wave Kernel Signature} (WKS). Yet, that approach does not reliably infers the signs of the antisymmetric eigenfunctions.


\subsection { Background}


\subsubsection {Laplace-Beltrami Eigendecomposition}
\label{sec:lbo}
Let us be given a shape modeled as a compact two-dimensional manifold $M$. 
The divergence of the gradient of a function $f$ over the manifold,
\begin{equation}
	\Delta_G f = \mathrm{div}\, \mathrm{grad} f
\label{eq:lbo_def},
\end{equation}
is called the \emph{Laplace-Beltrami operator (LBO)} of $f$ and can be considered as a generalization of the standard notion of the Laplace operator to manifolds \cite{taubin1995signal,levy2010spectral}.
The Laplace-Beltrami operator is completely derived from the  {\em metric tensor} $G$.
\begin{equation}
\label{eq:lbo_mt}
	\Delta_G f = \operatorname{div}\operatorname{grad} f = \underbrace{\frac{1}{\sqrt {|G|}} \sum\limits_{i} \partial_i \sqrt{|G|}}_{divergence} \underbrace{\sum\limits_{j}g^{ij} \partial_j}_{gradient} f \text ,
\end{equation}
where $g^{ij}=(G^{-1})_{ij}$ are the components of the inverse metric tensor.
Since the operator $- \Delta_G$ is a positive self-adjoint operator, it admits an eigendecomposition with non-negative eigenvalues $\lambda_i$ and corresponding orthonormal eigenfunctions $\phi_i$, 
\begin{equation}
\label{eq:helmoltz}
	-\Delta_G \phi_i = \lambda_i \phi_i \text,
\end{equation}
where orthonormality is understood in the sense of the local inner product induced by the Riemannian metric on the manifold.
Furthermore, due to the assumption that our manifold is compact, the spectrum is discrete.
We can order the eigenvalues as follows $0 = \lambda_1 < \lambda_2 < \cdots < \lambda_i < \cdots$ \quad .
The set of corresponding eigenfunctions given by $\{\phi_1, \phi_2, \cdots, \phi_i, \cdots\}$ forms an orthonormal basis of functions defined on $M$.
%


\subsubsection{Diffusion maps}
\label{subsec:diff_maps}
The heat equation describes the distribution of heat in time. On a manifold $M$, the heat equation is governed by the  Laplace-Beltrami operator $\Delta_G$,
\begin{equation}
	\frac{\partial u}{\partial t} = \Delta_G u \text .
\end{equation}
The heat kernel $K_t(x, y)$ is the diffusion kernel of the heat operator $e^{t\Delta_G} (t>0)$.
It is a fundamental solution of the heat equation with point heat source at $x$ (heat value at point $y$ after time $t$).
The heat kernel can be represented in the Laplace-Beltrami eigenbasis as
\begin{equation}
	K_t(x, y) = \sum\limits_i (\tilde{\lambda}_i)^t\phi_i(x)\phi(y) =  \sum\limits_i e^{-\lambda_i t}\phi_i(x)\phi_i(y) \text,
\end{equation}
where ${\tilde{\lambda}_i}$ are the eigenvalues of the heat operator, ${\lambda_i}$ are the eigenvalues of the LBO, and ${\tilde{\lambda}_i}=e^{-\lambda_i}$.

Using the  heat kernel we can define the diffusion distance \cite{coifman2005geometric}
\begin{equation}
\begin{split}
	d^2_{M,t}(x,y) & = ||K_t(x,\cdotp)-K_t(y,\cdotp)|| \\ & = \int_M(K_t(x,z)-K_t(y,z))^2 da(z) \text ,
\end{split}
\end{equation}
where $da$ is the area element of $M$.

The diffusion distance $d_{M,t}(x,y)$  can be computed by embedding the manifold into the infinite Euclidean space spanned by the LBO eigenbasis
\begin{equation}
\label{eq:spec_dist}
	d_{M,t}(x,y) = \bigg (\sum\limits_ie^{-2{\lambda}_i t}(\phi_i(x)-\phi_i(y))^2 \bigg )^\frac{1}{2} \text.
\end{equation}

The {\em diffusion map} $\{\Phi_t\}$ embeds the data into the finite $N$-dimension Euclidean space
\begin{equation}
	\Phi_t(x) = 
	\begin{bmatrix}
		e^{-{\lambda}_1 t}\phi_1(x) \\
	  	e^{-{\lambda}_2 t}\phi_2(x) \\
		 ... \\ 
		e^{-{\lambda}_N t}\phi_N(x) 
	\end{bmatrix} 
	 \text,
\end{equation}
  so that in this space, the
Euclidean distance is equal to the diffusion distance up to a relative truncation error
\begin{equation}
	d_{M,t}(x,y) \approx ||\Phi_t(x)-\Phi_t(y)|| \text .
\end{equation}


\subsubsection{Multivariate distribution comparison}
\label{subsec:mdc}
The distribution of $N$ continuous random variables $\phi_1,\phi_2,...,\phi_N$ is directly represented by the probability density function $f_{\phi_1,\phi_2,...,\phi_N}(\phi_1,\phi_2,...,\phi_N)$.
The direct estimation of the multivariate probability density function from data samples is hard to accomplish. Therefore, an indirect representation is often being utilized.
The probability distribution can be indirectly specified (under mild conditions) in a number of different ways, the simplest of which is by its raw moments
\begin{equation}
\label{eq:gmm_raw}
	\mu_{i_1,i_2,...,i_N} \equiv \operatorname{E}[\phi_1^{i_1}\phi_2^{i_2}...\phi_N^{i_N}], \quad \{i_1, i_2,...,i_N\} \in \mathbb Z_{\ge 0}.
\end{equation}

In order to compare the multivariate distributions of two sets of $N$ random variables $\phi^X_1,\phi^X_2,...,\phi^X_N$ and $\phi^Y_1,\phi^Y_2,...,\phi^Y_N$, we can use this indirect representation, and compare the raw moments of the random variables.
In practice, only a small set of the moments $\mathcal I$ can be used for measuring the difference between the distributions
\begin{equation}
\label{eq:dist_diff}
\begin{split}
	C_{X,Y} & = \sum\limits_{\{i_1,i_2,...,i_N\} \in \mathcal {I}}{\rho_{i_1,i_2,...,i_N}(\mu^X_{i_1,i_2,...,i_N}-\mu^Y_{i_1,i_2,...,i_N})^2},
\end{split}
\end{equation}
where $\rho_{i_1,i_2,...,i_N}$ are the weights associated with each raw moment.
%


\section {Eigenfunction matching}


\subsection {Problem formulation}
Let us denote by $X$ and $Y$ the two shapes we would like to match.
We represent the correspondence between $X$ and $Y$ by a bijective mapping $\varphi : X \mapsto Y$ , such that for each point $x \in X$, its corresponding point is $\varphi(x) \in Y$. 
The diffusion map embeds each point $x \in X$  into the $N$ dimension Euclidean space $\mathbb R^N$ according to  $\Phi_t^{X,N}(x)$.
Correspondingly, each point $y \in Y$ is embedded by the mapping $\Phi_t^{Y,N}(y)$ into $\mathbb R^N$. 
We denote the diffusion map at $t=0$ by $\Phi^{X}(x)=\Phi_{t=0}^{X,N}(x)$ and $\Phi^{Y}(y)=\Phi_{t=0}^{Y,N}(y)$, respectively.

We wish to find embeddings of shape $X$ and shape $Y$ to the finite dimensional Euclidean space, such that the corresponding points $x \in X$ and $\varphi(x) \in Y$ will be mapped to nearby points in the embedded space. 
Because of the inconsistencies described in the introduction, the diffusion maps of shapes $X$ and $Y$ do not necessarily fulfill this property. 
Our task is to modify the diffusion map  $\Phi^Y(y)$  by a small number of parameters $\theta$ such that the new embedding $\tilde{\Phi}_{\theta}^Y(y)$ will match $\Phi^X$, \ie $\Phi^{X}(x) \approx \tilde{\Phi}_{\theta}^{Y}(\varphi(x))$.

For the $N$ low eigenvalues the matching is characterized by the following parameters: 
\begin{itemize}
	\item The respective signs of the eigenfunctions $\textbf s: s_i \in \{+1, -1\}$.
	\item Permutation vector $\boldsymbol{\pi}$ of the eigenfunctions: $\pi : \{1,2, ...,N\} \mapsto \{1,2, ...,N\}$.
\end{itemize}
We would like to find the parameters $\hat\theta=\{\hat{\textbf s}; \hat{\boldsymbol{\pi}}\}$, that create the matched embedding $\tilde{\Phi}_{\hat\theta}^Y(y)$ with elements $\tilde{\phi}^Y_i=\hat{s}_i\phi^Y_{\hat\pi(i)},\quad i \in {1,2,...N}$.
%


\subsection {Matching cost function}
The entire algorithm can be expressed as the minimization of the following cost function
\begin{equation}
\label{eq:entire_cf}
\begin{split}
	\{\hat{\textbf s}; \hat{\boldsymbol{\pi}}\} = & \mbox{argmin}_{{\textbf s}; {\boldsymbol{\pi}}}(C(\textbf s, 
		\boldsymbol{\pi})+C^S(\textbf s, \boldsymbol{\pi}) \\ & +\alpha (C^P_\nabla(\textbf s, \boldsymbol{\pi})+C^{P,S}_\nabla(\textbf s, \boldsymbol{\pi})))\text.
\end {split}
\end{equation}

The terms of the cost function can be expressed by
\begin{itemize}
	\item $C(\textbf s, \boldsymbol{\pi}) = \sum\limits_{i,j,k}(\mu^X_{i,j,k} - s_i s_j s_k\mu^Y_{\pi(i),\pi(j),\pi(k)})^2$,

\quad \quad $\mu_{i,j,k} = \operatorname{E}[\phi_i\phi_j\phi_k], \quad i,j,k \in \{1,2, ...,N\}.$
	\item $C^P_\nabla(\textbf s, \boldsymbol{\pi}) = \sum\limits_{i,j,k,p}(\xi^X_{i,j,k,p}-s_i s_j s_k\xi^Y_{\pi(i),\pi(j),\pi(k),p})^2$,

		\quad \quad $\xi_{i,j,k,p} = \operatorname{E}[\nu_{i,j}\phi_k w_p(|\phi_k|)],$

		 \quad \quad $i,j,k \in \{1,2, ...,N\}, \quad p \in \{1..P\}$, \\

		\quad \quad $\nu_{i,j} = (\nabla_{G}\phi_i \times \nabla_{G}\phi_j) \cdot \textbf n$.

	\item $C^S(\textbf s, \boldsymbol{\pi}) = N\sum\limits_{i,q}(\mu^{X,S}_{i,q}-s_i\mu^{Y,S}_{\pi(i), q})^2$,

		\quad \quad $\mu^S_{i,q} = \operatorname{E}[\phi_i \psi_q], \quad i \in \{1,2, ...,N\}, \quad q \in \{1,2,...Q\}$.

	\item $C^{P,S}_\nabla(\textbf s, \boldsymbol{\pi}) = \sum\limits_{i,q,k,p}(\xi^{X,S}_{i,q,k,p}-s_i s_k\xi^{Y,S}_{\pi(i),q,\pi(k),p})^2$,

		\quad \quad  $\xi^S_{i,q,k,p} = \operatorname{E}[\nu^S_{i,q}\phi_k w_p(|\phi_k|)],$

		\quad \quad  $i,k \in \{1..N\}, p \in \{1..P\}, q \in \{1..Q\}$,

		\quad \quad $\nu^S_{i,q} = (\nabla_{G}\phi_i \times \nabla_{G}\psi_q) \cdot \textbf n$.
\end{itemize}

Where
\begin{itemize}
	\item $\phi_i$ are the eigenfunctions of the Laplace-Beltrami operator  $-\Delta_G\phi_i = \lambda_i \phi_i$.
	\item $w_p: \mathbb R_{\ge 0} \mapsto [0,1]$ are nonlinear weighting functions.
	\item $\psi_q:  M \mapsto \mathbb R$ are the components of an external point signature.
	\item $\nabla_{G}$ is the gradient induced by the metric tensor $G$.
	\item $\operatorname{E}[z] = \int_M{z da_M}$, where $da_M$ is the area element of the manifold $M$.
	\item $\textbf n$ is the normal to the surface.
	\item $\times$ is the cross product in $\mathbb R^3$ and $\cdot$ is the inner product in $\mathbb R^3$.
	\item The weighting parameter $\alpha$ determines the relative weight of the gradient cost functions.
\end{itemize}
In Appendix \ref{app:discretization} we give full details of the discretization we have used to implement the matching algorithm.
\\ \\
The application specific parameters include:
\begin{itemize}
\setlength{\itemindent}{1em}
	\item $N$ - The number of eigenfunctions to be matched.
	\item $\{w_p\}^P_{p=1}$  - The $P$ nonlinear weighting functions.
	\item $\{\psi_q\}^Q_{q=1}$  - The external point signature of size $Q$.
	\item $\alpha$ - The relative weight of the gradient cost functions.
\end{itemize}
In Appendix \ref{app:specific_params} we give the details of the application specific parameters that were used in our experiments.
\\ \\
Next, we review the different terms of the cost function.
%


\subsubsection{Resolving sign ambiguities and permutations}
\label{sec:sgn_amb}

For now, let us limit our discussion to resolving the sign ambiguity $\textbf s$.
If we had known the correspondence between the two shapes, the sign of the $ i_{th}$ eigenfunction $s_i$ could be inferred by pointwise comparison
\begin{equation}
	\hat s_i = \mbox{argmin}_{s_i}\operatorname{E}[(\phi^X_i(x)-s_i\phi^Y_i(\varphi(x))^2] \text ,
\end{equation}
and the expectation is taken over the manifold
\begin{equation}
	\operatorname{E}(f(x)) = \int_X f(x) da_X,
\end{equation}
where $da_X$ is the area element of the shape $X$.
Unfortunately, the correspondence is unknown. Hence,  pointwise comparison cannot be used in a straightforward manner.

We now make the analogy between the values of the eigenfunctions over the manifold and $N$ random variables.
We consider the vector of values of the diffusion map $\Phi^{X}(x)$ at point $x$ as a sample out of a multivariate distribution $f_{\Phi}(\phi_1(x),\phi_2(x),...,\phi_N(x))$. 
We wish to match the multivariate distributions $f_{\Phi^X}$ and $f_{\Phi^Y_{\theta}}$. As explained in Section \ref{subsec:mdc}, an indirect representation of the distribution is suitable for comparing multivariate distributions. 
Specifically, we shall use the raw moments.

By way of construction, the non-trivial eigenfunctions have zero mean and are orthonormal. Hence, the first and second moments carry no information.
Accordingly, we must use higher order moments to match the distributions. We propose to use the third order moments over the manifold $M$
\begin{equation}
\begin{split}
	\mu_{i,j,k} & = \operatorname{E}[\phi_i\phi_j\phi_k] = \int_M\phi_i\phi_j\phi_k da_M, \\ & \quad i,j,k \in \{1,2, ...,N\}.
\end{split}
\end{equation}


\subsubsection{Resolving antisymmetric eigenfunctions}
For shapes with {\em intrinsic symmetries} (see \cite{raviv2007symmetries}) some of the eigenfunctions have antisymmetric distributions. 
The distribution of the antisymmetric eigenfunctions is agnostic to sign change.
Hence, the signs of the antisymmetric eigenfunctions cannot be resolved by the simple scheme described in section \ref{sec:sgn_amb}.

The gradient of the eigenfunctions $\nabla \phi_k$ could be exploited to resolve the sign ambiguity:
\begin{itemize}
	\item The gradient $\nabla f$ of an antisymmetric eigenfunction $f$ is not antisymmetric.
	\item The gradient is a linear operator. Consequently $\nabla (-f) = -\nabla f, \quad \forall f$.
\end{itemize}
Therefore, we can farther expand the set of variables that are used in the calculation of the raw moments, by incorporating the gradient.
The gradient vector is contained in the tangent plane.
Thus, the cross product of the gradients of two eigenfunctions points either outward or inward of an orientable surface.
Changing the sign of one eigenfunction will flip the direction of the cross product.
We can use this property to define new functions $\nu_{i,j}$ over the manifold
\begin{equation}
\label{eq:nu_cross}
	\nu_{i,j} = (\nabla\phi_i \times \nabla\phi_j) \cdot \textbf n,
\end{equation}
where $\textbf n$ is the outward pointing normal to the tangent plane.
We shall use the joint moments of the eigenfunctions and their gradients
\begin{equation}
	\xi_{i,j,k} = \operatorname{E}[\nu_{i,j}\phi_k], \quad i,j,k \in \{1,2, ...,N\}.
\end{equation}
We note that Equation (\ref{eq:nu_cross}) can be farther simplified by
\begin{equation}
	\nu_{i,j} = (\nabla\phi_i \times \nabla\phi_j) \cdot \textbf n = \nabla\phi_i \cdot (\nabla\phi_j \times \textbf n).
\end{equation}
$(\nabla\phi_j \times \textbf n)$ can be computed only once for each $\phi_j$.

\subsubsection{Raw moments over regions}
Taking the expectation over the whole shape may be too crude, especially for detecting antisymmetric sign ambiguities.
We can refine the minimization criterion by taking the expectation over different regions.
Remember that the correspondence between the shapes is yet unknown, therefore, directly dividing the shape into corresponding regions is impossible.
Indirectly dividing the shape to different regions is possible by using the eigenfunctions themselves.
The eigenfunctions $\phi_k$, $k \in \{1...N\}$ have respective low eigenvalues which means that they have a slow rate of change.
Therefore, it is possible to define functions $w_p(|\phi_k|)$, $p \in \{1..P\}$ in a way that will output high or low values at different regions. 
For example, we can define $h(|\phi_k|) = 1\text{ if }|\phi_k|>\text{TH} \text{ and zero otherwise}$, where $\text{TH}$ is a scalar threshold.
The output of these functions automatically divides the two shapes in a similar manner, without the use of pointwise correspondence.
Moreover, because the function $w_p(|\phi_k|)$ is symmetric, its output does not depend on the sign of the eigenfunction $\phi_k$.
We conclude that we can use $w_p(|\phi_k|)$ to make a weighted average of the raw moments according to different regions
\begin{equation}
\begin{split}
\xi_{i,j,k,p} & = \operatorname{E}[\nu_{i,j}\phi_k w_p(|\phi_k|)], \\ & \quad i,j,k \in \{1,2, ...,N\}\quad p \in \{1..P\}.
\end{split}
\end{equation}


\subsubsection{Pointwise signatures as side information}
We can easily use other signatures $(\psi_1, \psi_2, ..., \psi_Q)$ as side information to refine the minimization criterion.
Specifically, we can use signatures that carry no inconsistencies among different shapes.
In our experiments we used the {\em Heat Kernel Signature}  (HKS) as an additional signature \cite{sun2009concise}.
We can use the joint moments of the diffusion maps and the additional signatures $\psi_q$
\begin{equation}
\label{eq:third_order}
	\mu^S_{i,q} = \operatorname{E}[\phi_i \psi_q], \quad i \in \{1,2, ...,N\}, \quad q \in \{1,2,...Q\},
\end{equation}
and compute the cross of the eigenfunctions gradient $\nabla\phi_i$ and the signature functions gradients $\nabla\psi_q$
\begin{equation}
	\nu^S_{i,q} = (\nabla\phi_i \times \nabla\psi_q) \cdot \textbf n,
\end{equation}
\begin{equation}
\begin{split}
	\xi^S_{i,q,k,p} & = \operatorname{E}[\nu^S_{i,q}\phi_k w_p(|\phi_k|)], \\ &  i,k \in \{1,2, ...,N\}\quad p \in \{1..P\} \quad q \in \{1..Q\}.
\end{split}
\end{equation}


\subsection {Solving the minimization problem}
The minimization of Equation (\ref{eq:entire_cf})  is a non-convex optimization problem. 
Yet, it only involves a small number of discrete parameters.
Therefore, an exhaustive search is possible.
In practice, we implemented the search in four steps:
\begin{itemize}
	\item 	Step 1 - An initialization of $\textbf s^{\textbf 0}$ is determined by $s_i = \operatorname{sign}(\mu^X_{i,i,i}\mu^Y_{i,i,i})$ and $\boldsymbol{\pi^0} = [0,1,...N]$.
	\item 	Step 2 - The permutation vector $\hat{\boldsymbol{\pi}}$ is found by minimizing $C(\textbf s, \boldsymbol{\pi})+C^S(\textbf s, \boldsymbol{\pi})$. 
	We make an educated guess for the possible permutations, limiting the search for two permutation profiles:
	\begin{itemize}
	\item[$\diamond$] 	two consecutive eigenfunction switching (with possible sign change), \ie $[\pi_i,\pi_j,\pi_k,\pi_l]=[j,i,l,k], \quad j=i+1,  l=k+1$
	\item[$\diamond$] 	triplet permutation (with possible sign change), \ie $[\pi_i,\pi_j,\pi_k]=[j,k,i] $ or $[k,i,j], \quad j=k+1, i=j+1$;
	\end {itemize}
	\item Step 3: The sign sequence is resolved again by minimizing $C(\textbf s, \boldsymbol{\pi})+C^S(\textbf s, \boldsymbol{\pi})$. 
			In this step {\em all} possible quadruple sign changes are checked, setting the permutation vector found in Step 2.
			If the cost function was decreased in Step 2 or Step 3, then return to Step 2.
			While finding the optimal sign sequence and permutation vector, we keep a list of all possible good sign sequences for the next step.
	\item 	Step 4: The optimal sign sequence $\hat{\textbf s}$ is found by comparing the entire cost function $C(\textbf s, \boldsymbol{\pi})+C^S(\textbf s, \boldsymbol{\pi})+\alpha (C^P_\nabla(\textbf s, \boldsymbol{\pi})+C^{P,S}_\nabla(\textbf s, \boldsymbol{\pi}))$ for each sign sequence in the list created in Step 3.
\end{itemize}

We note that the computation of the moments can be done prior to the minimization algorithm. This calculation of the raw moments of shape $X$ is independent of shape $Y$ and can be performed for each shape before the matching procedure.
The entire cost function given in Equation (\ref{eq:entire_cf}) can be computed from the raw moments and the parameters, without the use of the eigenfunctions themselves.
%


\section{Results}
\label{sec:results}

\begin{figure}[t]
	\begin{center}
	\begin{overpic}[width=1.1\columnwidth]{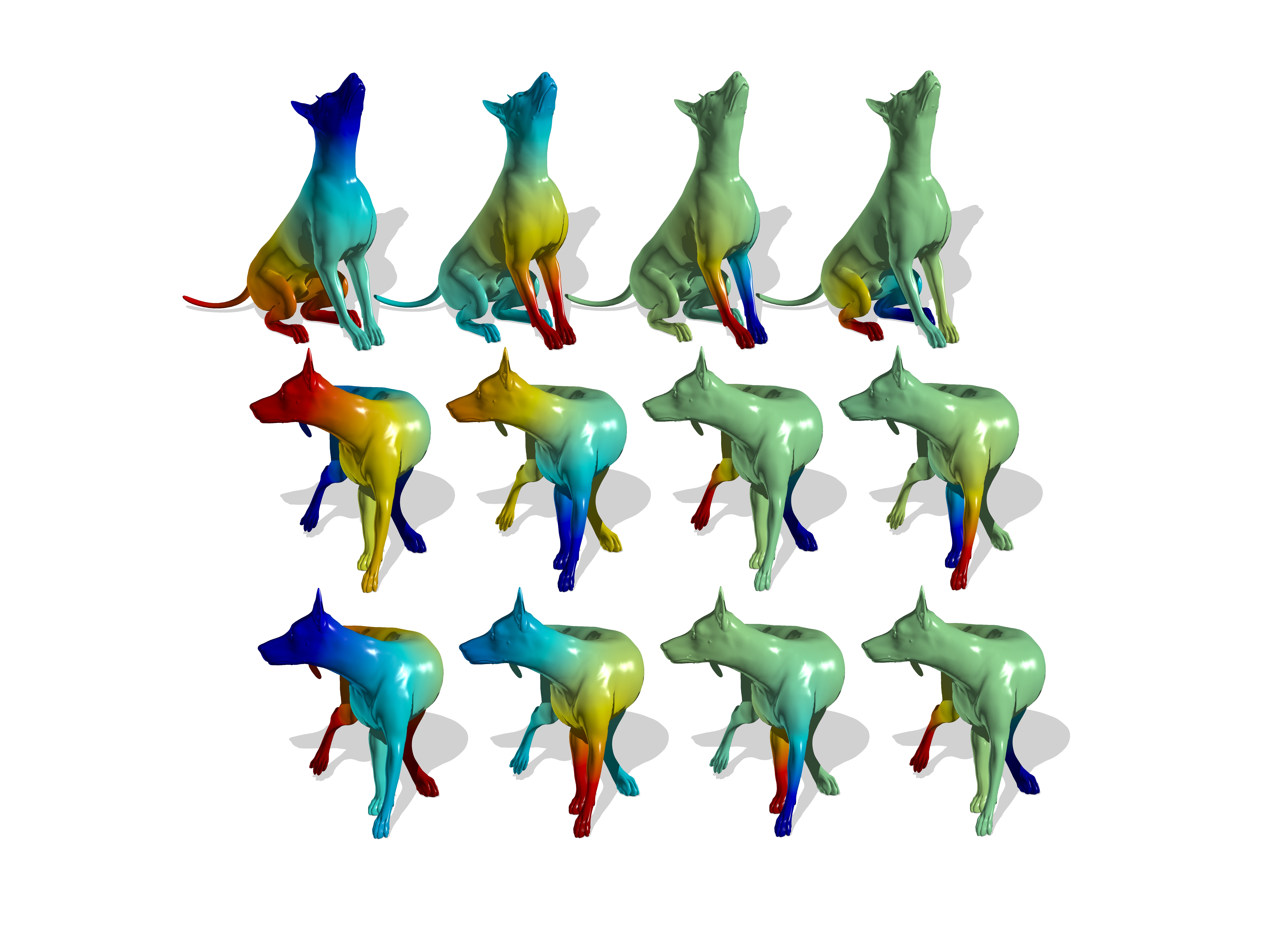}
		\put(10,60){\Large$\Phi^X$}
		\put(10,40){\Large$\Phi^Y$}
		\put(10,21){\Large$\tilde{\Phi}^Y$}
		\put(25,5){\Large${\phi_1}$}
		\put(40,5){\Large${\phi_2}$}
		\put(55,5){\Large${\phi_3}$}
		\put(70,5){\Large${\phi_4}$}
	\end{overpic}
	\end{center}
	\caption{\small Eigenfunction matching of two nearly isometric shapes. Hot and cold colors represent positive and negative values, respectively. Top: first pose of a dog. Center: second pose of a dog. Bottom: second pose of a dog after matching algorithm.}
	\label{fig:dog1_dog10_ef}
\end{figure}
\begin{figure}[t]
	\begin{center}
	\begin{overpic}[width=0.9\columnwidth]{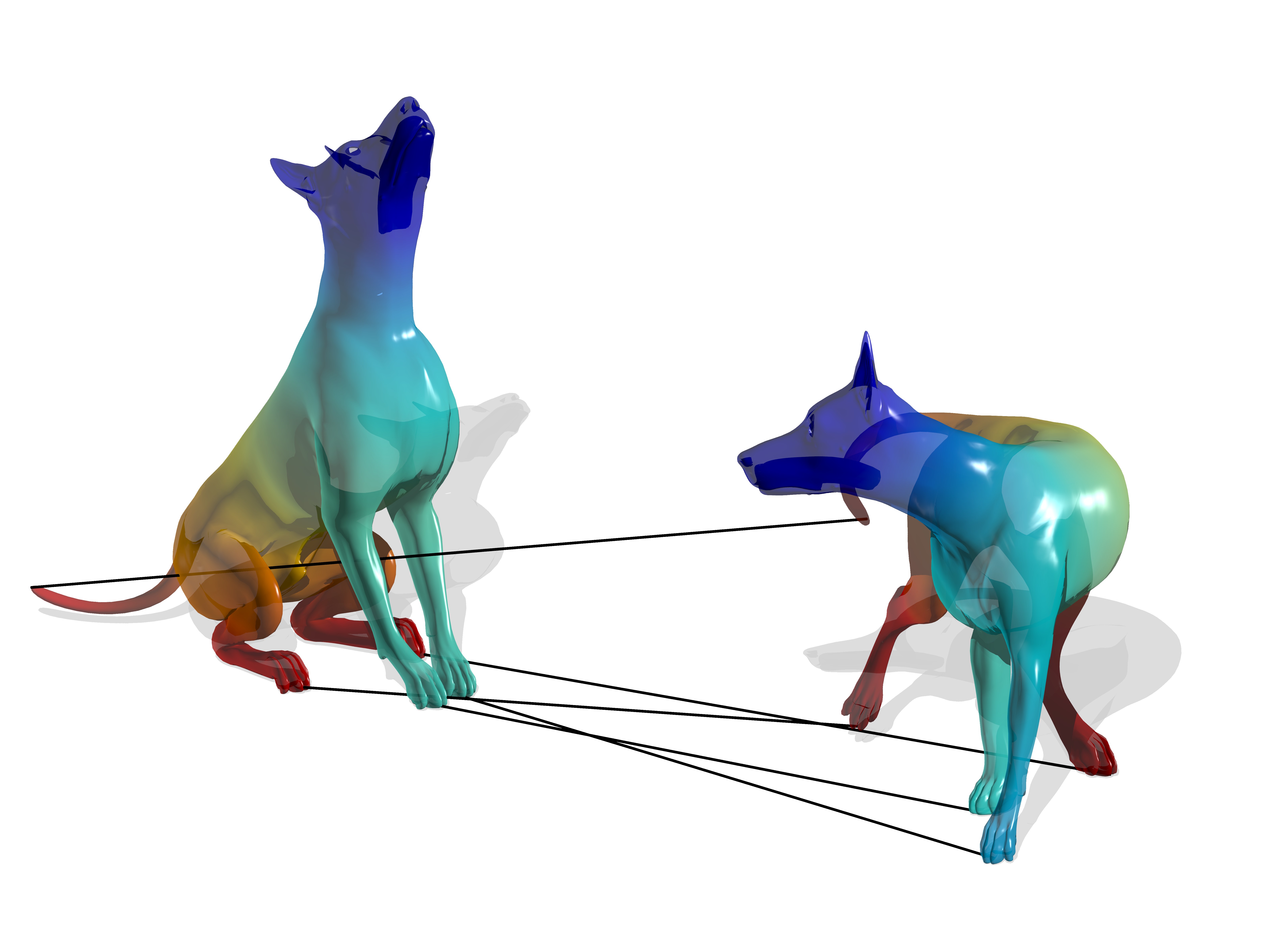}
	\end{overpic}
	\end{center}
	\caption{\small Feature point correspondence of two nearly isometric shapes of a horse.}
	\label{fig:dog1_dog10_fp}
\end{figure}

\begin{figure}[t]
	\begin{center}
	\begin{overpic}[width=1.1\columnwidth]{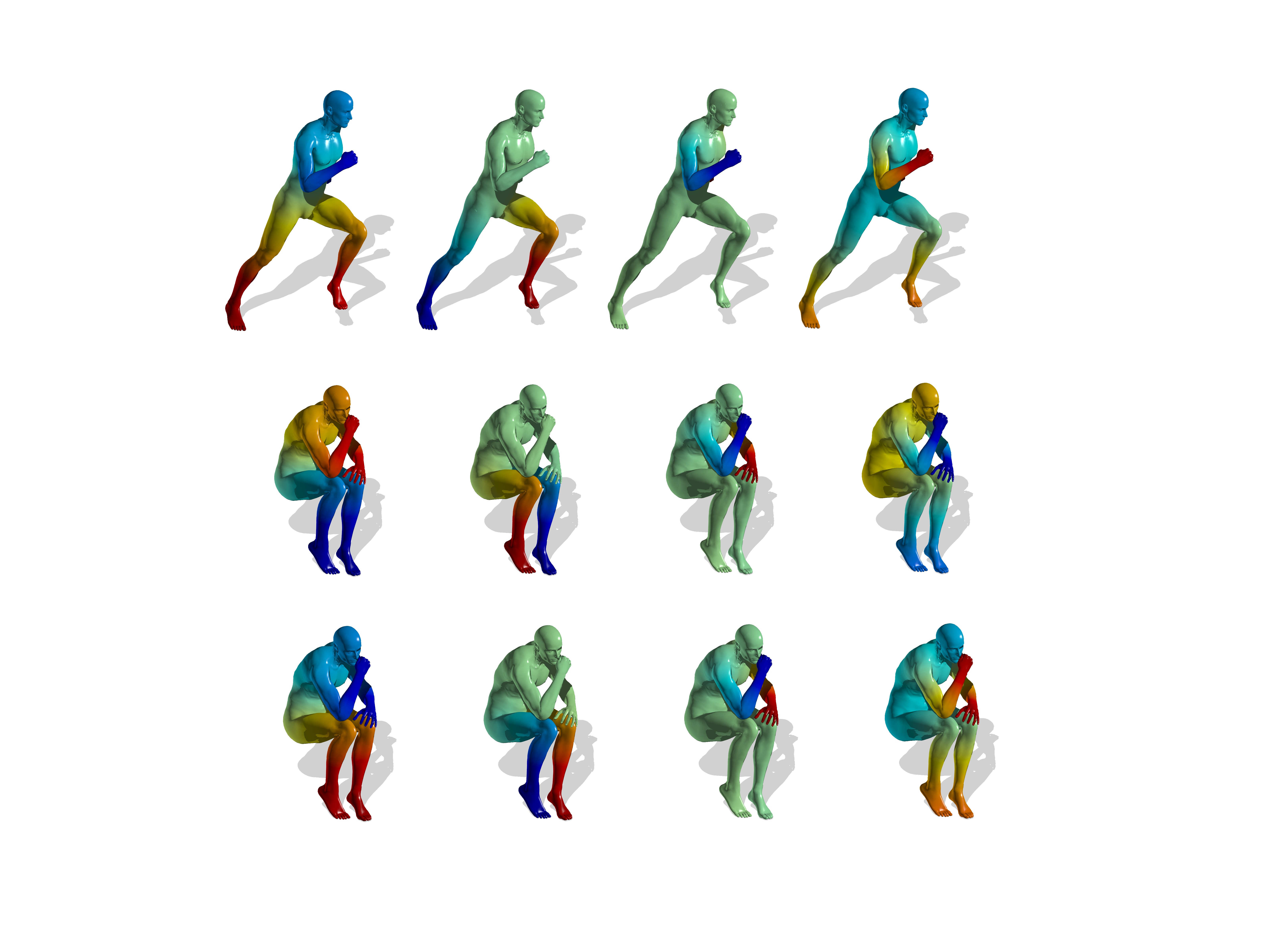}
		\put(10,60){\Large$\Phi^X$}
		\put(10,40){\Large$\Phi^Y$}
		\put(10,21){\Large$\tilde{\Phi}^Y$}
		\put(25,5){\Large${\phi_1}$}
		\put(40,5){\Large${\phi_2}$}
		\put(55,5){\Large${\phi_3}$}
		\put(70,5){\Large${\phi_4}$}
	\end{overpic}
	\end{center}
	\caption{\small Eigenfunction matching of two nearly isometric shapes. Hot and cold colors represent positive and negative values, respectively. Top: first pose of a human. Center: second pose of a human. Bottom: second pose of a human after matching algorithm.}
	\label{fig:michael16_michael14_ef}
\end{figure}
\begin{figure}[t]
	\begin{center}
	\begin{overpic}[width=0.9\columnwidth]{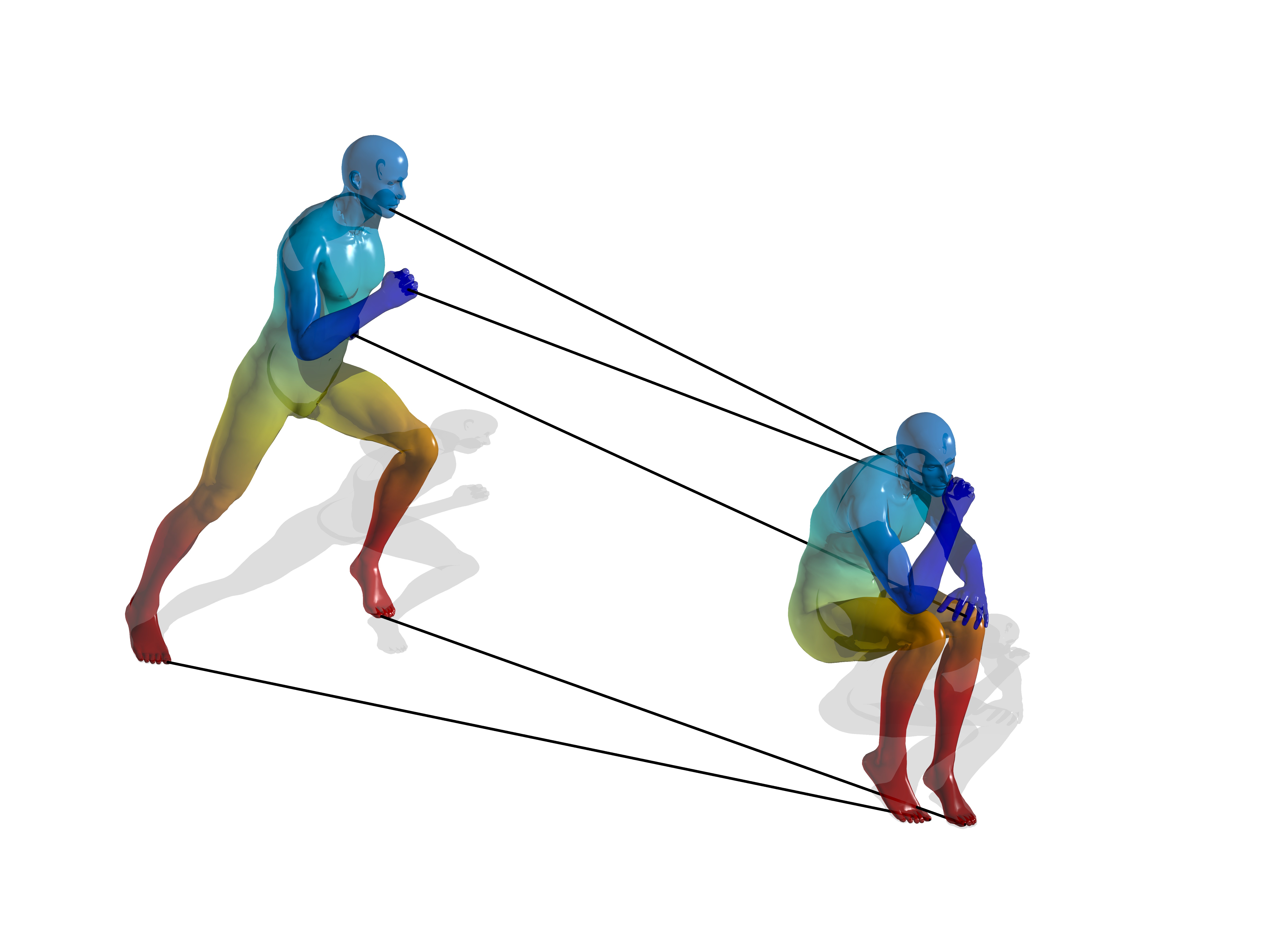}
	\end{overpic}
	\end{center}
	\caption{\small Feature point correspondence of two nearly isometric shapes of a horse.}
	\label{fig:michael16_michael14_fp}
\end{figure}

\begin{figure}[t]
	\begin{center}
	\begin{overpic}[width=1.1\columnwidth]{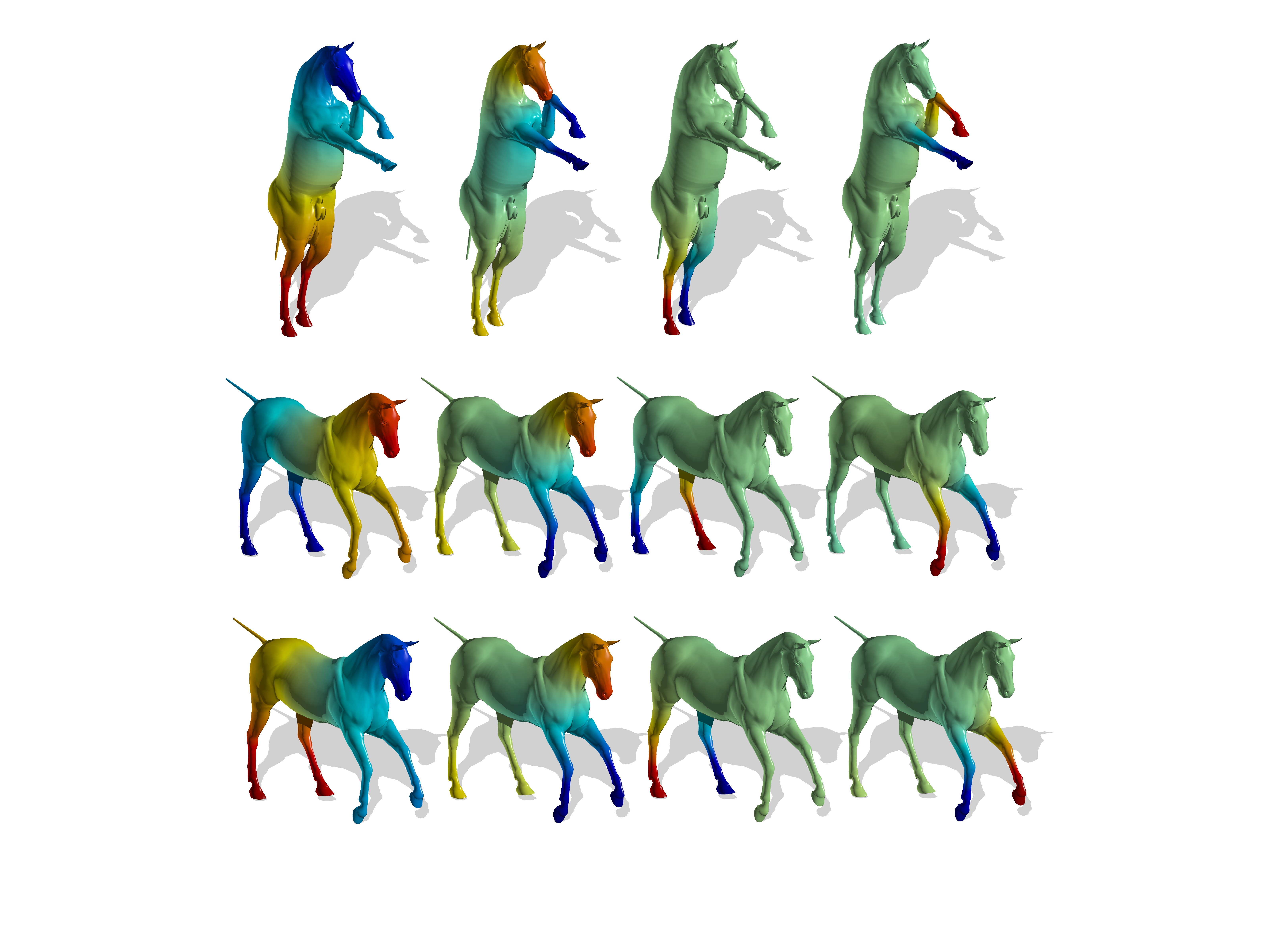}
		\put(10,60){\Large$\Phi^X$}
		\put(10,40){\Large$\Phi^Y$}
		\put(10,21){\Large$\tilde{\Phi}^Y$}
		\put(25,5){\Large${\phi_1}$}
		\put(40,5){\Large${\phi_2}$}
		\put(55,5){\Large${\phi_3}$}
		\put(70,5){\Large${\phi_4}$}
	\end{overpic}
	\end{center}
	\caption{\small Eigenfunction matching of two nearly isometric shapes. Hot and cold colors represent positive and negative values, respectively. Top: first pose of a horse. Center: second pose of a horse. Bottom: second pose of a horse after matching algorithm.}
	\label{fig:horse15_horse17_ef}
\end{figure}
\begin{figure}[t]
	\begin{center}
	\begin{overpic}[width=0.9\columnwidth]{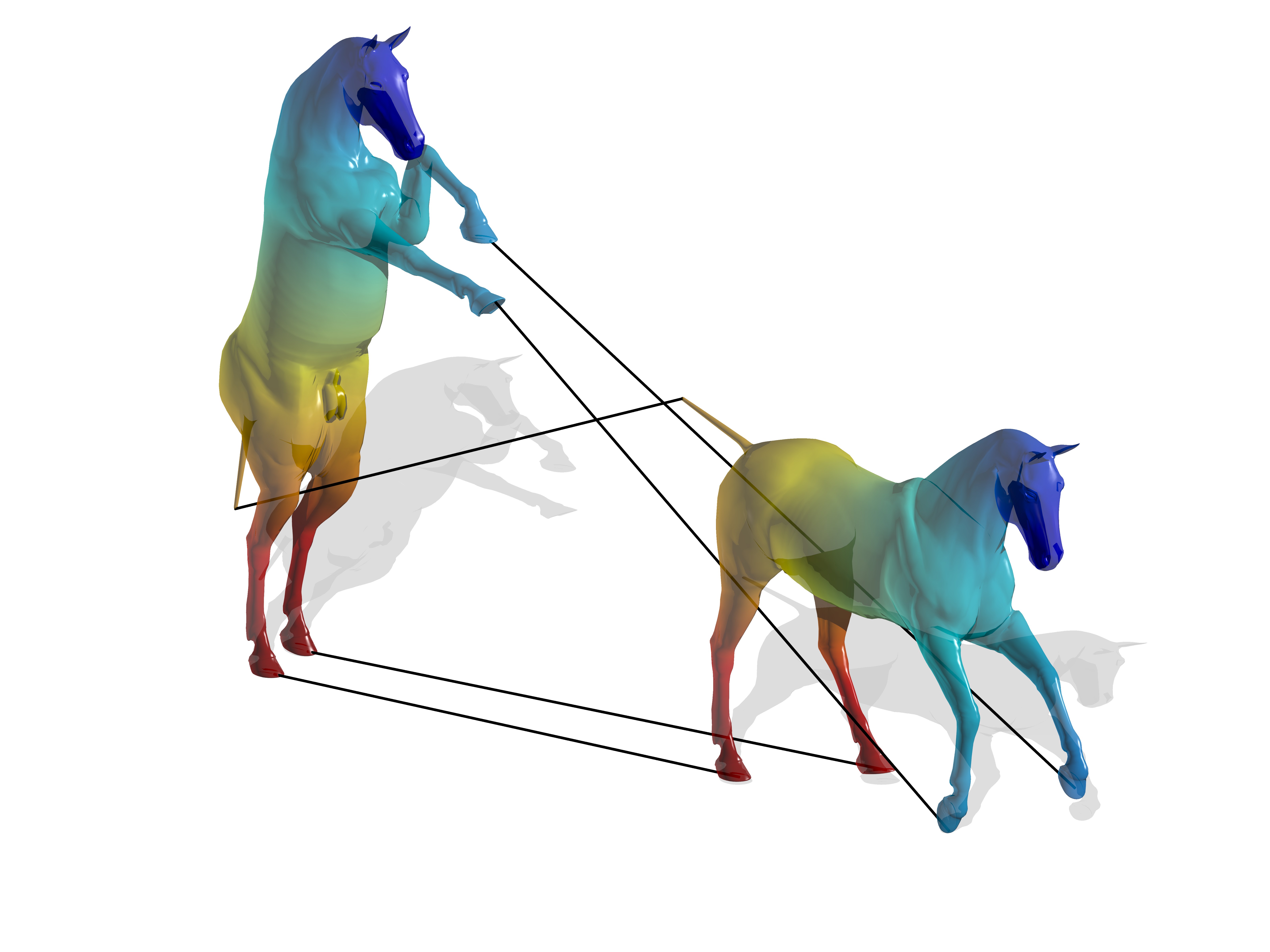}
	\end{overpic}
	\end{center}
	\caption{\small Feature point correspondence of two nearly isometric shapes of a horse.}
	\label{fig:horse15_horse17_fp}
\end{figure}

We tested the proposed method on pairs of shapes represented by triangulated meshes from the TOSCA  database \cite{bronstein2008numerical}.
Figures \ref{fig:dog1_dog10_ef}, \ref{fig:michael16_michael14_ef} and \ref{fig:horse15_horse17_ef} show how the proposed method succeeds in matching the eigenfunctions of several isometric shapes.
In each figure, at the top, are the first four eigenfunctions of the first pose of the object. In the middle, the eigenfunctions of the second pose of the object. 
At the bottom, are the first four eigenfunctions of the second pose with the correct sign sequence and permutations. 

For example, we can see in Figure \ref{fig:dog1_dog10_ef} that the eigenfunction matching algorithm swapped $\phi^Y_3$ and $\phi^Y_4$. It also correctly flipped the signs of $\phi^Y_1$, $\phi^Y_2$ and $\phi^Y_4$. We also notice that the matching algorithm was able to detect the correct signs of the antisymmetric eigenfunctions. For example, in Figure \ref{fig:michael16_michael14_ef}, the sign of the antisymmetric eigenfunction $\phi^Y_2$ was correctly flipped, while keeping the sign of $\phi^Y_3$.

We applied the matched eigenfunctions for detecting feature point correspondence between the two shapes. 
A selected number of feature points from the first shape were matched to the second one using a combination of two signatures:
\begin{itemize}
	\item The matched low order eigenfunctions that represent global strucure of the shapes.
	\item The {\em Heat Kernel Signature} (HKS) derivative (Equation (\ref{eq:hksd})), that being a bandpass filter, expresses more local features.
\end{itemize}
Figures \ref{fig:dog1_dog10_fp}, \ref{fig:michael16_michael14_fp} and \ref{fig:horse15_horse17_fp} show that the correspondences between feature points were found correctly. Notice that this approach was able to resolve the symmetries of the given shapes.

\section{Conclusion}
The  Laplace Beltrami operator (LBO) provides us with a flat eigenspace in which surfaces could be represented as canonical forms in an isometric invariant manner.
However, the order and directions (signs) of the axises in this Hilbert space do not have to correspond when two isometric surfaces are considered. 
In order to resolve such potential ambiguities we resorted to high order statistics  of the eigenfunctions of the LBO and their interaction with the surface normal.
It appears that these cross moments allow for ordered directional matching of the components of corresponding eigenspaces.
We demonstrated that resolving the sign and order correspondence allows for shape matching in various scenarios. 
In the future we plan on extending the proposed framework to enable it to deal with more generic transformations, like the scale invariant metric introduced by Aflalo \etal \cite{aflalo2013scale}.



\subsection*{Acknowledgments}
We thank Anastasia Dubrovina  for helpful discussions and suggestions during the progress of this work.


\appendix

\section{Discretization}
\label{app:discretization}

\subsection{Laplace-Beltrami eigendecomposition}
We used the cotangent weight scheme for the Laplace–Beltrami operator discretization, proposed by Pinkall \etal \cite{Pinkall93computingdiscrete} and later refined by Meyer \etal \cite{meyer2002discrete}.
In order to calculate the eigendecomposition of the Laplace–Beltrami operator we solved the generalized eigendecomposition problem, as suggested by Rustamov \cite{rustamov2007laplace}. 
\begin{eqnarray}
\label{eq:lbo_dis}
	W\phi = \lambda A \phi \text,
\end{eqnarray}
where
\begin{eqnarray}
	w_{ij} = \begin{cases} \frac{\cot\alpha_{ij} + \cot\beta_{ij}}{2} & i\neq j \\ \sum\limits_{k\ne i} w_{ik} & i=j \end{cases} \text,
\end{eqnarray}
and $A$ is a diagonal matrix. $A_{ii}$ equals the voroni area about vertex $i$.

\subsection{Gradient}
We assume that the function $f$ is linear over the triangle with vertices $V_i, V_j, V_k$ with values $f_i, f_j, f_k$ at the vertices. 
We define the local coordinates $(u,v)$ with coordinates $(0,0)$, $(0,1)$, $(1,0)$ at the vertices $V_i, V_j, V_k$.
Because $f$ is assumed to be linear 
\begin{eqnarray}
	\frac{\partial f}{\partial u} = f_j-f_i
\end{eqnarray}
 and 
\begin{eqnarray}
	\frac{\partial f}{\partial v} = f_k-f_i \text,
\end{eqnarray}
which can be written as 
\begin{eqnarray}
	\frac{\partial f}{\partial (u,v)} = \begin{bmatrix} \frac{\partial f}{\partial u} \\[0.3em] \frac{\partial f}{\partial v} \end{bmatrix}^T = (DF)^T,
\end{eqnarray}
where $D =\begin{bmatrix} -1 & 1 & 0\\ -1 & 0 & 1\end{bmatrix}$ and $F =\begin{bmatrix} f_i \\ f_j \\ f_k \end{bmatrix}$. 

The Jacobian 
\begin{eqnarray}
J=\frac{\partial(x,y,z)}{\partial(u,v)} = \begin{bmatrix} \frac{\partial x}{\partial u} & \frac{\partial x}{\partial v} \\[0.3em] \frac{\partial y}{\partial u} & \frac{\partial y}{\partial v} \\[0.3em] \frac{\partial z}{\partial u} & \frac{\partial z}{\partial v} \end{bmatrix}^T = [V_j-V_i, V_k-V_i]^T.
\end{eqnarray}
By the chain rule $\frac{\partial f}{\partial(u,v)} = \frac{\partial f}{\partial (x,y,z)}\frac{\partial(x,y,z)}{\partial(u,v)}$ and in matrix form
\begin{eqnarray}
	(DF)^T = (\nabla f)^TJ^T,
\end{eqnarray}
or equivalently
\begin{eqnarray}
	J\nabla f = DF \text .
\end{eqnarray}
By taking the pseudoinverse we get the discrete gradient operator over a triangle 
\begin{eqnarray}
\label{eq:grad_dis}
	\nabla =  J^T(JJ^T)^{-1} D
\end{eqnarray}


\section{Application specific parameters}
\label{app:specific_params}
In our experiments we used the following specific parameters for the matching algorithm:
\begin{itemize}
\setlength{\itemindent}{1em}
	\item We matched $N=10$ eigenfunctions.
	\item Soft thresholding was used to define $P = 2$ nonlinear weighting functions $w_p$:
	\begin{equation}
	\begin{split}
	& w_0(z) = \begin{cases} 0, & \mbox{if } |z|<\mbox{TH} \\ 1, & \mbox{if } |z|>2\mbox{TH} \\ (|z|-\mbox{TH})/\mbox{TH} & \mbox{otherwise} \quad \quad , \end{cases}\\
	& w_1(z) = 1-w_0(z) ,
	\end{split} 
	\end{equation}
	where $\mbox{TH} = 0.1\frac{1}{\sqrt{\int_Mda}}$.
	\item For generating the external pointwise signature $\psi_q$, the Heat Kernel Signature (HKS) was used \cite{sun2009concise}. In the approximation of the Heat Kernel Signature $\text{HKS}_t(x) = \sum\limits_{i=1}^h e^{-\lambda_it}\phi^2_i(x)$, we used $h=30$ eigenfunctions. We used a bandpass filter form of the HKS by taking the derivative of the Heat Kernel Signature. The HKS derivative  was logarithmically sampled $Q=6$ times at $t=t_q, \quad q={1,2,...Q}$, with $t_1=\frac{1}{50\lambda_1}$ and $t_Q = \frac{1}{\lambda_1}$. $\psi_q$ were normalized according to the inner product over the manifold.
	\begin{equation}
	\label{eq:hksd}
		\begin{split}
		& \psi_q(x) = \frac{\tilde{\psi}_q(x)}{\sqrt{\int_M\tilde{\psi}^2_q(\tilde{x})da(\tilde{x})}}, \\
		& \tilde{\psi}_q(x) = \frac{\partial}{\partial t}\text{HKS}_{t}(x) \quad \mbox{sampled at} \quad t=t_q, \\
		& \frac{\partial}{\partial t}\text{HKS}_{t}(x) = \sum^h_{i=1}-\lambda_i e^{-\lambda_it}\phi^2_i(x).
		\end{split}
	\end{equation}
	\item The relative weight parameter $\alpha$ was set by balancing the influence of the terms of the cost function.	\begin{equation}
		\alpha = \frac{\sum\limits_{i,j,k}(\mu^X_{i,j,k})^2 + N\sum\limits_{i,q}(\mu^{X,S}_{i,q})^2}{\sum\limits_{i,j,k,p}(\xi^X_{i,j,k,p})^2 + \sum\limits_{i,q,k,p}(\xi^{X,S}_{i,q,k,p})^2}.
	\end{equation}
\end{itemize}

\bibliographystyle{alpha}
\bibliography{Matching-LBO-eigenspace}

\end{document}